\documentclass{aa}  

\usepackage{graphicx}
\usepackage{amsmath,amsfonts,amssymb}
\usepackage{natbib}

\usepackage{txfonts}
\usepackage{xcolor}

\usepackage{blindtext}
\usepackage{float}
\usepackage{dblfloatfix}
\usepackage{afterpage}
\usepackage{ifthen}
\usepackage[morefloats=12]{morefloats}

\usepackage{placeins}
\usepackage{multicol}
\bibpunct{(}{)}{;}{a}{}{,}
\usepackage[switch]{lineno}
\definecolor{linkcolor}{rgb}{0.6,0,0}
\definecolor{citecolor}{rgb}{0,0,0.75}
\definecolor{urlcolor}{rgb}{0.12,0.46,0.7}
\usepackage[breaklinks, colorlinks, urlcolor=urlcolor,
    linkcolor=linkcolor,citecolor=citecolor,pdfencoding=auto]{hyperref}
\hypersetup{linktocpage}
\usepackage{bold-extra}

\usepackage[capitalise]{cleveref}
\crefname{section}{Sect.}{Sects.}
\crefname{table}{Table}{Tables}
\crefname{equation}{Eq.}{Eqs.}
\crefname{appendix}{Appendix}{Appendices}
\Crefname{section}{Section}{Sections}
\Crefname{table}{Table}{Tables}
\Crefname{equation}{Equation}{Equations}
\crefname{appendix}{Appendix}{Appendices}

\makeatletter
\AddToHook{cmd/appendix/before}{\def\cref@section@alias{appendix}}
\makeatother

\def\setsymbol#1#2{\expandafter\def\csname #1\endcsname{#2}}
\def\getsymbol#1{\csname #1\endcsname}

\def\Planck{\textit{Planck}}

\newbox\tablebox    \newdimen\tablewidth
\def\leaderfil{\leaders\hbox to 5pt{\hss.\hss}\hfil}
\def\endPlancktable{\tablewidth=\columnwidth 
    $$\hss\copy\tablebox\hss$$
    \vskip-\lastskip\vskip -2pt}

\def\tablenote#1 #2\par{\begingroup \parindent=0.8em
    \abovedisplayshortskip=0pt\belowdisplayshortskip=0pt
    \noindent
    $$\hss\vbox{\hsize\tablewidth \hangindent=\parindent \hangafter=1 \noindent
    \hbox to \parindent{$^#1$\hss}\strut#2\strut\par}\hss$$
    \endgroup}
\def\doubleline{\vskip 3pt\hrule \vskip 1.5pt \hrule \vskip 5pt}

\def\L2{\ifmmode L_2\else $L_2$\fi}

\def\DeltaT{\ifmmode \Delta T\else $\Delta T$\fi}
\def\deltat{\ifmmode \Delta t\else $\Delta t$\fi}
\def\fknee{\ifmmode f_{\rm knee}\else $f_{\rm knee}$\fi}
\def\Fmax{\ifmmode F_{\rm max}\else $F_{\rm max}$\fi}
\def\solar{\ifmmode{\rm M}_{\mathord\odot}\else${\rm M}_{\mathord\odot}$\fi}
\def\Msolar{\ifmmode{\rm M}_{\mathord\odot}\else${\rm M}_{\mathord\odot}$\fi}
\def\Lsolar{\ifmmode{\rm L}_{\mathord\odot}\else${\rm L}_{\mathord\odot}$\fi}
\def\inv{\ifmmode^{-1}\else$^{-1}$\fi}
\def\mo{\ifmmode^{-1}\else$^{-1}$\fi}
\def\sup#1{\ifmmode ^{\rm #1}\else $^{\rm #1}$\fi}
\def\expo#1{\ifmmode \times 10^{#1}\else $\times 10^{#1}$\fi}
\def\,{\thinspace}
\def\lsim{\mathrel{\raise .4ex\hbox{\rlap{$<$}\lower 1.2ex\hbox{$\sim$}}}}
\def\gsim{\mathrel{\raise .4ex\hbox{\rlap{$>$}\lower 1.2ex\hbox{$\sim$}}}}

\def\simprop{\mathrel{\raise .4ex\hbox{\rlap{$\propto$}\lower 1.2ex\hbox{$\sim$}}}}
\def\deg{\ifmmode^\circ\else$^\circ$\fi}
\def\pdeg{\ifmmode $\setbox0=\hbox{$^{\circ}$}\rlap{\hskip.11\wd0 .}$^{\circ}
          \else \setbox0=\hbox{$^{\circ}$}\rlap{\hskip.11\wd0 .}$^{\circ}$\fi}
\def\arcs{\ifmmode {^{\scriptstyle\prime\prime}}
          \else $^{\scriptstyle\prime\prime}$\fi}
\def\arcm{\ifmmode {^{\scriptstyle\prime}}
          \else $^{\scriptstyle\prime}$\fi}
\newdimen\sa  \newdimen\sb
\def\parcs{\sa=.07em \sb=.03em
     \ifmmode \hbox{\rlap{.}}^{\scriptstyle\prime\kern -\sb\prime}\hbox{\kern -\sa}
     \else \rlap{.}$^{\scriptstyle\prime\kern -\sb\prime}$\kern -\sa\fi}
\def\parcm{\sa=.08em \sb=.03em
     \ifmmode \hbox{\rlap{.}\kern\sa}^{\scriptstyle\prime}\hbox{\kern-\sb}
     \else \rlap{.}\kern\sa$^{\scriptstyle\prime}$\kern-\sb\fi}
\def\ra[#1 #2 #3.#4]{#1\sup{h}#2\sup{m}#3\sup{s}\llap.#4}
\def\dec[#1 #2 #3.#4]{#1\deg#2\arcm#3\arcs\llap.#4}
\def\deco[#1 #2 #3]{#1\deg#2\arcm#3\arcs}
\def\rra[#1 #2]{#1\sup{h}#2\sup{m}}

\def\dots{\relax\ifmmode \ldots\else $\ldots$\fi}
\def\WHzsr{\ifmmode $W\,Hz\mo\,sr\mo$\else W\,Hz\mo\,sr\mo\fi}
\def\mHz{\ifmmode $\,mHz$\else \,mHz\fi}
\def\GHz{\ifmmode $\,GHz$\else \,GHz\fi}
\def\mKs{\ifmmode $\,mK\,s$^{1/2}\else \,mK\,s$^{1/2}$\fi}
\def\muKs{\ifmmode \,\mu$K\,s$^{1/2}\else \,$\mu$K\,s$^{1/2}$\fi}
\def\muKRJs{\ifmmode \,\mu$K$_{\rm RJ}$\,s$^{1/2}\else \,$\mu$K$_{\rm RJ}$\,s$^{1/2}$\fi}
\def\muKHz{\ifmmode \,\mu$K\,Hz$^{-1/2}\else \,$\mu$K\,Hz$^{-1/2}$\fi}
\def\MJysr{\ifmmode \,$MJy\,sr\mo$\else \,MJy\,sr\mo\fi}
\def\MJysrmK{\ifmmode \,$MJy\,sr\mo$\,mK$_{\rm CMB}\mo\else \,MJy\,sr\mo\,mK$_{\rm CMB}\mo$\fi}
\def\microns{\ifmmode \,\mu$m$\else \,$\mu$m\fi}

\def\muK{\ifmmode \,\mu$K$\else \,$\mu$\hbox{K}\fi}
\def\microK{\ifmmode \,\mu$K$\else \,$\mu$\hbox{K}\fi}
\def\muW{\ifmmode \,\mu$W$\else \,$\mu$\hbox{W}\fi}
\def\kms{\ifmmode $\,km\,s$^{-1}\else \,km\,s$^{-1}$\fi}
\def\kmsMpc{\ifmmode $\,\kms\,Mpc\mo$\else \,\kms\,Mpc\mo\fi}
\providecommand{\sorthelp}[1]{}

\newcommand{\mathsc}[1]{{\normalfont\textsc{#1}}}
\newcommand{\dv}[0]{\vec{d}}

\newcommand{\N}[0]{\tens{N}}

\renewcommand{\a}[0]{\vec{a}}
\newcommand{\n}[0]{\vec{n}}
\renewcommand{\t}[0]{\vec{t}}
\def\Cosmoglobe{\textsc{Cosmoglobe}}
\def\Planck{\textit{Planck}}

\def\COBE{\textit{COBE}}
\def\GAIA{\textit{Gaia}}

\def\Gaia{\textit{Gaia}}

\def\IRAS{\textit{{IRAS}}}
\def\nside{$N_{\mathrm{side}}$}

\newcommand{\CII}{\ensuremath{\mathsc{C\ ii}}}

\newcommand{\HI}{\ensuremath{\mathsc{H\ i}}}
\def\Commander{\texttt{Commander} }
\def\commander{\texttt{Commander}}

\def\Tcmb{\ifmmode T_\mathrm{CMB}\else $T_{\mathrm{CMB}}$\fi}
\def\Tcold{\ifmmode T_\mathrm{c}\else $T_{\mathrm{c}}$\fi}
\def\Thot{\ifmmode T_\mathrm{h}\else $T_{\mathrm{h}}$\fi}
\def\Tnear{\ifmmode T_\mathrm{n}\else $T_{\mathrm{n}}$\fi}
\def\scmb{\ifmmode s_\mathrm{CMB}\else $s_{\mathrm{CMB}}$\fi}
\def\squad{\ifmmode s_\mathrm{quad}\else $s_{\mathrm{quad}}$\fi}
\def\ssynch{\ifmmode s_\mathrm{s}\else $s_\mathrm{s}$\fi}
\def\sdust{\ifmmode s_\mathrm{d}\else $s_{\mathrm{d}}$\fi}
\def\ssdust{\ifmmode s_\mathrm{sd}\else $s_{\mathrm{sd}}$\fi}
\def\same{\ifmmode s_\mathrm{AME}\else $s_{\mathrm{AME}}$\fi}
\def\ssrc{\ifmmode s_\mathrm{src}\else $s_{\mathrm{src}}$\fi}
\def\sco{\ifmmode s_\mathrm{CO}\else $s_{\mathrm{CO}}$\fi}
\def\sff{\ifmmode s_\mathrm{ff}\else $s_{\mathrm{ff}}$\fi}
\def\gff{\ifmmode g_\mathrm{ff}\else $g_{\mathrm{ff}}$\fi}
\def\fsynch{\ifmmode f_\mathrm{s}\else $f_{\mathrm{s}}$\fi}
\def\fsd{\ifmmode f_\mathrm{sd}\else $f_{\mathrm{sd}}$\fi}
\def\fame{\ifmmode f_\mathrm{AME}\else $f_{\mathrm{AME}}$\fi}
\def\alphasrc{\ifmmode \alpha_\mathrm{src}\else $\alpha_{\mathrm{src}}$\fi}
\def\bcold{\ifmmode \beta_\mathrm{c}\else $\beta_{\mathrm{c}}$\fi}
\def\bhot{\ifmmode \beta_\mathrm{h}\else $\beta_{\mathrm{h}}$\fi}
\def\bnear{\ifmmode \beta_\mathrm{n}\else $\beta_{\mathrm{n}}$\fi}
\def\bsynch{\ifmmode \beta_\mathrm{s}\else $\beta_{\mathrm{s}}$\fi} 
\def\bsun{\ifmmode \beta_\mathrm{sun}\else $\beta_{\mathrm{sun}}$\fi} 
\def\nuzeros{\ifmmode \nu_{0,\mathrm{s}}\else $\nu_{0,\mathrm{s}}$\fi} 
\def\nuzeroff{\ifmmode \nu_{0,\mathrm{ff}}\else $\nu_{0,\mathrm{ff}}$\fi} 
\def\nuzerocold{\ifmmode \nu_{0,\mathrm{c}}\else $\nu_{0,\mathrm{c}}$\fi}
\def\nuzerohot{\ifmmode \nu_{0,\mathrm{h}}\else $\nu_{0,\mathrm{h}}$\fi}
\def\nuzeronear{\ifmmode \nu_{0,\mathrm{n}}\else $\nu_{0,\mathrm{n}}$\fi} 
\def\nuzeroame{\ifmmode \nu_{0,\mathrm{AME}}\else $\nu_{0,\mathrm{AME}}$\fi} 
\def\nuzerosd{\ifmmode \nu_{0,\mathrm{}}\else $\nu_{0,\mathrm{sd}}$\fi} 
\def\nuzerosrc{\ifmmode \nu_{0,\mathrm{src}}\else $\nu_{0,\mathrm{src}}$\fi} 
\def\nup{\ifmmode \nu_{\mathrm{p}}\else $\nu_{\mathrm{p}}$\fi} 
\def\alphasd{\ifmmode \alpha_{\mathrm{sd}}\else $\alpha_{\mathrm{sd}}$\fi} 
\def\Te{\ifmmode T_{\mathrm{e}}\else $T_{\mathrm{e}}$\fi} 
\def\kB{\ifmmode k_\mathrm{B}\else $k_{\mathrm{B}}$\fi}

\newcommand{\x}{\checkmark}

\begin{document} 
%\linenumbers

%\title{\bfseries{\Cosmoglobe\ DR2. On the correlation between ionized carbon and thermal dust emission}}
%\title{\bfseries{\Cosmoglobe\ DR2. Thermal dust and ionized carbon emission are spatially strongly correlated}}
\title{\bfseries{\Cosmoglobe\ DR2. V. Spatial correlations between thermal dust and ionized carbon emission in \Planck\ HFI and COBE-DIRBE}}

   \newcommand{\oslo}[0]{1}
\newcommand{\milan}[0]{2}
\newcommand{\ijclab}[0]{3}
\newcommand{\gothenberg}[0]{4}
\newcommand{\trento}[0]{5}
\newcommand{\milanoinfn}[0]{6}
\author{\small
E.~Gjerl\o w\inst{\oslo}\thanks{Corresponding author: E.~Gjerløw; \url{eirik.gjerlow@astro.uio.no}}
\and
R.~M.~Sullivan\inst{\oslo}
\and
R.~Aurvik\inst{\oslo}
\and
A.~Basyrov\inst{\oslo}
\and
L.~A.~Bianchi\inst{\oslo}
\and
A.~Bonato\inst{\milan}
\and
M.~Brilenkov\inst{\oslo}
\and
H.~K.~Eriksen\inst{\oslo}
\and
U.~Fuskeland\inst{\oslo}
\and
M.~Galloway\inst{\oslo}
\and
K.~A.~Glasscock\inst{\oslo}
\and
L.~T.~Hergt\inst{\ijclab}
\and
D.~Herman\inst{\oslo}
\and
J.~G.~S.~Lunde\inst{\oslo}
\and
M.~San\inst{\oslo}
\and
A.~I.~Silva Martins\inst{\oslo}
\and
D.~Sponseller\inst{\gothenberg}
\and
N.-O.~Stutzer\inst{\oslo}
\and
H.~Thommesen\inst{\oslo}
\and
V.~Vikenes\inst{\oslo}
\and
D.~J.~Watts\inst{\oslo}
\and
I.~K.~Wehus\inst{\oslo}
\and
L.~Zapelli\inst{\milan,\trento,\milanoinfn}
}
\institute{\small
Institute of Theoretical Astrophysics, University of Oslo, Blindern, Oslo, Norway\goodbreak
\and
Dipartimento di Fisica, Università degli Studi di Milano, Via Celoria, 16, Milano, Italy
\and
Laboratoire de Physique des 2 infinis -- Irène Joliot Curie (IJCLab), Orsay, France
\and
Department of Space, Earth and Environment, Chalmers University of Technology, Gothenburg, Sweden\goodbreak
\and
Università di Trento, Università degli Studi di Milano, CUP E66E23000110001\goodbreak
\and
INFN sezione di Milano, 20133 Milano, Italy\goodbreak
}

   % Shortened title, author list for top of page 
   \titlerunning{Correlations between thermal dust and \CII}
   \authorrunning{Gjerløw et al.}

   \date{\today} 
   
   \abstract{We fit five tracers of thermal dust emission to ten \textit{Planck} HFI and \textit{COBE}-DIRBE frequency maps between 353\,GHz and 25\,THz, aiming to map the relative importance of each physical host environment as a function of frequency and position on the sky. Four of these correspond to classic thermal dust tracers, namely \textsc{H i} (HI4PI), CO \citep{dame2001}, H$\alpha$ (WHAM, \cite{wham:2003,2016WHAM}), and dust extinction (\textit{Gaia}; \citealp{edenhofer:2024}), while the fifth is ionized carbon (\textsc{C ii}) emission as observed by \textit{COBE}-FIRAS. To our knowledge, this has until now been considered to be primarily a gas tracer, rather than a dust tracer. After smoothing all data to the common resolution of the FIRAS experiment, and subtracting subdominant astrophysical components as appropriate for each channel (cosmic microwave and infrared backgrounds, and zodiacal light), we jointly fit these five templates to each frequency channel through standard multi-variate linear regression. At frequencies higher than 1\,THz, we find that the dominant tracer is in fact \textsc{C ii}, and above 10\,THz this component accounts for almost the entire fitted signal; at frequencies below 1\,THz, its importance is second only to \textsc{H i}. We further find that all five components are well described by a modified blackbody spectral energy density (SED) up to some component-dependent maximum frequency ranging between 1 and 5\,THz. In this interpretation, the \textsc{C ii}-correlated component is the hottest among all five, with an effective temperature of about 25\,K. For comparison, the \textsc{H i} and CO components have effective temperatures of 16\,K and 12\,K, respectively. The H$\alpha$ component has a temperature of 18\,K, and, unlike the other four, is observed in absorption rather than emission. The spectral indices of the five components range between $\beta = 1.4$ (for \textsc{H i}) and 2.6 (for H$\alpha$); for the \textsc{C ii} component, $\beta=1.56$. Despite the simplicity of this model, which relies only on external templates coupled to spatially isotropic SEDs, we find that it captures 98\,\% of the full signal root mean squared (RMS) below 1\,THz. At higher frequencies, which are more susceptible to non-thermal emission processes, the model still captures more than 80\,\% of the full signal RMS. This high efficiency suggests that spatial variations in the thermal dust SED, as for instance reported by \Planck\ and other experiments, may be more economically modelled on large angular scales in terms of a spatial mixing of individually isotropic physical components, than by a single uniform well-mixed interstellar medium coupled to a spatially varying temperature field, as has been the norm until now. Indeed, the results found in this paper motivate the thermal dust model adopted for the \textsc{Cosmoglobe} DR2 re-analysis of the \textit{COBE}-DIRBE data, and we believe that they may also provide inspiration for refining both current theoretical interstellar medium models and component separation algorithms in general.}
   

   \maketitle

%\setcounter{tocdepth}{2}
%\tableofcontents

% INTRODUCTION
%-------------------------------------------------------------------
\section{Introduction}
%\the\textwidth \the\columnwidth

The interstellar medium (ISM) plays a ubiquitous role in modern
astrophysics and cosmology across the electromagnetic spectrum (e.g., \citealt{draine2011}, \citealt{Hensley2023}). On 
one hand, understanding the composition and physics of the ISM informs
us about the structure and dynamics of both the Milky Way and distant
galaxies, and ISM studies are therefore an important and interesting
field of astronomy in their own right. On the other hand, ISM radiation
is a key contaminant for a broad range of other high-impact
science targets, for instance the search for gravitational waves in
the cosmic microwave background (CMB), dark matter annihilation in
gamma-ray observations, or dark energy constraints through
high-precision measurements of distant supernovae. Accurate ISM
modelling is therefore a key aspect in 21st century
astrophysics.

Broadly speaking, the ISM is comprised of cosmic rays
(relativistic particles), gas (atoms or small molecules), and dust
(large molecules, typically ranging in size from a few angstroms to
100\,$\mu$m). All of these emit electromagnetic radiation at various
frequencies, for instance through synchrotron, bremsstrahlung, quantum
mechanical line emission, or thermal emission. In addition, dust
grains absorb electromagnetic radiation with wavelengths that are
comparable to the grain size, which due to the grain distribution
happens most notably in wavelengths ranging from infrared to
X-rays. 

The current paper is part of a suite of seven companion papers that
describes the \Cosmoglobe\ Data Release 2 (DR2; \citealp{CG02_01}). The
primary main goal of this work is to reanalyze the \COBE-DIRBE data
\citep{hauser1998} within a global Bayesian end-to-end analysis
framework, and use the resulting data to constrain the spectrum of the
cosmic infrared background (CIB; \citealp{CG02_03}). The DIRBE
instrument observed the full sky in 10 wavelength bands covering 1.25
and 240\,$\mu$m. By virtue of covering virtually the entire infrared
regime, DIRBE is an excellent dust tracer, both originating from the
Milky Way and from within the Solar system. Indeed, while DIRBE's
original science goal was to detect and characterize the CIB spectrum
and fluctuations, the longest lasting legacy of the survey has
arguably been to serve as a unique well-calibrated full-sky tracer of
dust emission in the Milky Way 
\citep{schlegel1998,IRIS,planck2014-a12,sano2016,pysm_methods}
%(ADD LOTS OF REFERENCES)
and zodiacal
light emission \citep[e.g.,][]{K98,planck2013-pip88,2016AJ....151...71K,CG02_02,obrien2025}.% (include other recent references, CIBER for example).

One of the many important questions in which the DIRBE data has played a role
regards the nature and composition of thermal dust emission in the
far-infrared regime, and how it may be modelled most
efficiently. Early studies quickly suggested that a single so-called
modified blackbody (MBB) spectrum provided a good fit to the thermal
dust spectral energy density (SED) for a broad range of frequencies
\citep{reach1995b}. An MBB spectrum has three degrees of freedom, namely
an overall amplitude (tracing the density of the medium), an effective
temperature, and a spectral index. To this date, the
single-component MBB model plays a key role in both microwave
modelling and infrared-based dust studies.

However, it also quickly became clear that the validity of a single
MBB spectrum is limited in frequencies. Notably, in a seminal paper \citet{finkbeiner:1999} presented a two-component
MBB thermal dust model derived from the \IRAS\ 100\,$\mu$m and DIRBE 100
and 240\,$\mu$m data that served as a benchmark for the CMB community for more than a decade, and was
only superseded by the far more sensitive \Planck\ HFI data
\citep{planck2013-p06b}. Since that time, the combination of
\Planck, DIRBE, and \IRAS\ measurements have dominated the study of
large-scale dust emission at microwave and far-infrared frequencies.

\begin{figure}
  \centering
  \includegraphics[width=\columnwidth]{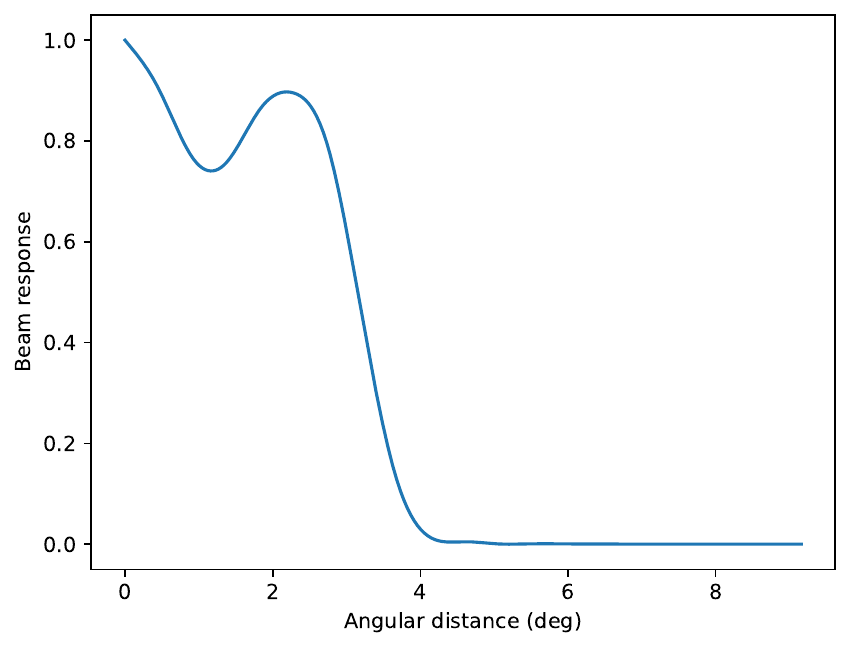}
  \caption{The \COBE-FIRAS beam, with which all maps in this analysis were smoothed. The beam diameter is around $7^\circ$ on the sky, which limits this analysis to \nside=16.}
  \label{fig:firasbeam}
\end{figure}

Despite massive efforts developing increasingly more accurate
and well-defined models of thermal dust emission, combining
theoretical insights with observational constraints \citep[e.g.,][]{draine2011,Hensley2023},
many critical questions plague this field to date. Furthermore,
the importance of understanding microwave dust emission has only increased in recent
years, as the attention of the CMB field has shifted to the search for
inflationary gravitational waves through $B$-mode polarization
constraints \citep[e.g.,][]{Bicep2018limit,SO2019,litebird2022}. For such experiments, polarized thermal dust emission
represents one of the most important contaminants, and major resources
are currently being invested in devising both instrumentation and
data analysis techniques to handle this challenge. 

% \begin{table}
%     \centering
%     \caption{Templates used for each frequency band.}
%     \begin{tabular}{c|c|c|c|c|c}
%         \label{tab:bands}
%         Band & \CII & HI4PI & Gaia & WHAM & CO \\
%         \hline
%         \Planck\ 353\,GHz & x & x & x & x & x \\
%         \Planck\ 545\,GHz & x & x & x & x & x \\
%         \Planck\,857 GHz & x & x & x & x & x \\
%         DIRBE 240 $\mu m$ & x & x & x & x & x \\
%         DIRBE 140 $\mu m$ & x & x & x & x & x \\
%         DIRBE 100 $\mu m$ & x & x & x & x & x \\
%         DIRBE 60 $\mu m$ & x & x & x & x & \\
%         DIRBE 25 $\mu m$ & x & x & & & \\
%         DIRBE 12 $\mu m$ & x & & & & 
%     \end{tabular} 
% \end{table}

In the current paper, and in two companion papers by \citet{CG02_06}
and \citet{CG02_07}, we revisit the question of how to model thermal
dust emission efficiently in light of the currently reprocessed DIRBE
data, jointly with archival measurements from
\Planck\ HFI. Specifically, during the course of the \Cosmoglobe\ DR2
work, a new four-component MBB model, based on five astrophysical tracers, has emerged as a particularly
compact and efficient description. Although a four-component model may
at first sight appear as significantly more complicated than either
the standard single-component MBB model adopted by the \Planck\ team,
or by the two-component MBB model pioneered by
\citet{finkbeiner:1999}, the key point in our new analysis is that all
four components are well traced by well-known dust templates coupled
with spatially isotropic SEDs. The number of degrees of freedom per
pixel is therefore very small, allowing for a very rigid and compact
statistical description.

Four of the five templates in this model are already known to be
efficient dust tracers, namely 1) HI4PI \HI\ line emission, tracing
molecular gas; 2) \GAIA\ extinction, tracing near-by dust absorption;
3) \citet{dame:2001} CO line emission, tracing cold dust in
star-forming regions; and 4) WHAM H$\alpha$ line emission, tracing hot
dust in ionized regions. However, in this paper and its companion
papers, we additionally find that ionized carbon emission, as traced
by the FIRAS \CII\ 158\,$\mu$m line emission map, is an excellent
tracer of thermal dust emission with a temperature of
$\sim$25-30\,K. This observation may have far-reaching consequences
for future studies and models of dust emission in the microwave and
infrared regimes.

In the current paper, we perform a linear regression analysis of
\Planck\ HFI 353--857\,GHz and DIRBE 3.5--100\,$\mu$m data with this
model, which demonstrates the validity of this key finding, even with a
minimal set of assumptions. In a follow-up analysis by \citet{CG02_06},
we perform a Bayesian analysis of \Planck\ HFI data with the goal of
deriving high-resolution templates of the \HI\ and \CII\ correlated
components directly from microwave data. Finally, \citet{CG02_07}
applies this novel model to the \COBE-DIRBE data.

\begin{table}
 \caption{Templates used for each frequency band.}
 \label{tab:bands}
\begingroup
\newdimen\tblskip \tblskip=4pt
\nointerlineskip
\vskip -3mm
\footnotesize
\setbox\tablebox=\vbox{
\halign{
 \tabskip 0pt \hbox to 0.25\linewidth{#\hfil}
&\hbox to 0.15\linewidth{\hfil#\hfil}\tabskip 0pt
&\hbox to 0.11\linewidth{\hfil#\hfil}\tabskip 0pt
&\hbox to 0.11\linewidth{\hfil#\hfil}\tabskip 0pt
&\hbox to 0.11\linewidth{\hfil#\hfil}\tabskip 0pt
&\hbox to 0.11\linewidth{\hfil#\hfil}\tabskip 0pt
&\hbox to 0.11\linewidth{\hfil#}\tabskip 0pt\cr
\noalign{\doubleline\vskip 1pt}
\omit\hbox to 1in{Band\hfil} & Freq. (GHz) & \CII & HI4PI & \Gaia & WHAM & CO \cr
\noalign{\vskip 4pt\hrule\vskip 6pt}
\noalign{\vskip 3pt}
       \Planck\ 353\,GHz & 353 & \x & \x & \x & \x & \x \cr
       \noalign{\vskip 3pt}
       \Planck\ 545\,GHz & 545 & \x & \x & \x & \x & \x \cr
       \noalign{\vskip 3pt}
       \Planck\,857 GHz & 857  & \x & \x & \x & \x & \x \cr
       \noalign{\vskip 3pt}
       DIRBE 240 $\mu$m & 1250 & \x & \x & \x & \x & \x \cr
       \noalign{\vskip 3pt}
       DIRBE 140 $\mu$m & 2100 & \x & \x & \x & \x & \x \cr
       \noalign{\vskip 3pt}
       DIRBE 100 $\mu$m & 3000 & \x & \x & \x & \x & \x \cr
       \noalign{\vskip 3pt}
       DIRBE 60 $\mu$m & 5000  & \x & \x & \x & \x & \cr
       \noalign{\vskip 3pt}
       DIRBE 25 $\mu$m & 12000 & \x & \x & & & \cr
       \noalign{\vskip 3pt}
       DIRBE 12 $\mu$m & 25000 & \x & & & & \cr
       \noalign{\vskip 3pt}
\noalign{\vskip 3pt\hrule\vskip 4pt}}}
\endPlancktable
\endgroup
\end{table}

The rest of the paper is organized as follows:
%In \cref{sec:dr2} we give a brief overview of the data model adopted for the\Cosmoglobe\ DR2 reanalysis. 
in \cref{sec:tempfit} we discuss the template fitting algorithm and in \cref{sec:data} the data used in the
current paper; finally, in \cref{sec:results} we present the results from
these calculations, before we conclude in \cref{sec:conclusions}.

\section{Template fitting methodology}
\label{sec:tempfit}

The main goal of this paper is to quantify the efficiency of known dust tracers for modelling thermal dust emission in the microwave and infrared frequency ranges. While all other papers in the current \Cosmoglobe\ DR2 paper suite adopts a global Bayesian analysis framework as their computational engine (as implemented in the \commander\ code; \citealp{eriksen:2004,BP03}), the analyses considered in this paper can be performed with much simpler template fitting tools. In addition to resulting in a much shorter computational run time, we also consider it valuable that the following results may be reproduced by other readers with a minimum of coding efforts and independently of \commander.

For each dataset\footnote{Bold symbols indicate a vector of
$n_\mathrm{pix}$ HEALPix pixels.} $\dv_i$ at a given frequency band
$i$, our basic data model reads
\begin{equation}
    \dv_i = \sum_c a_{i, c} \t_c + m_i + \n_i,
    \label{eq:datamodel}
\end{equation}
where the sum over $c$ runs over $n_{\mathrm{comp}}=5$ dust
components, each with an amplitude $a_{i, c}$ and a spatial template
$\t_c$. In addition, we include a monopole term, $m_i$, and a white
noise contribution, $\n_i$. We assume that all data are provided as
HEALPix\footnote{\url{http://healpix.sourceforge.net/}}
\citep{healpix,Zonca2019} maps with identical angular and pixel
resolution, and that the frequency maps only include dust emission and
noise; other astrophysical components, such as CMB, CIB, or zodiacal
light contributions are removed through preprocessing; see
\cref{sec:data} for details.

\begin{figure*}
  \centering
  \includegraphics[width=0.49\textwidth]{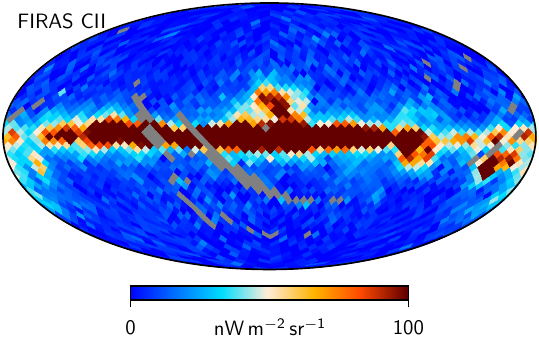}
  \includegraphics[width=0.49\textwidth]{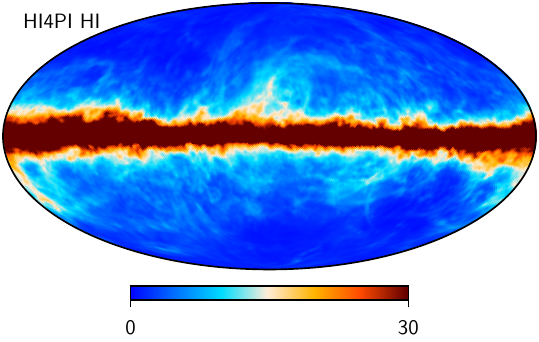}\\
  \includegraphics[width=0.49\textwidth]{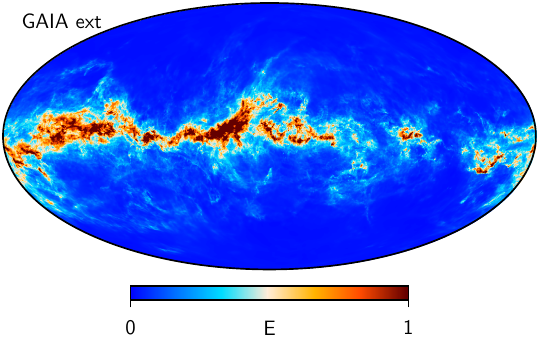}
  \includegraphics[width=0.49\textwidth]{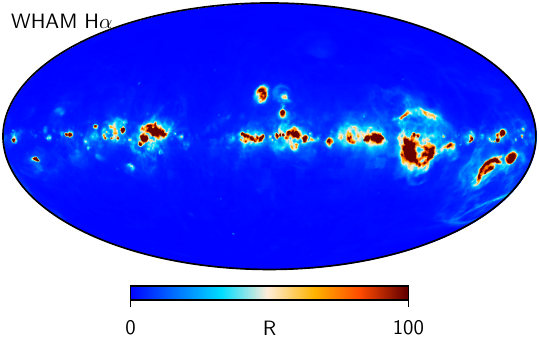}\\
  \includegraphics[width=0.49\textwidth]{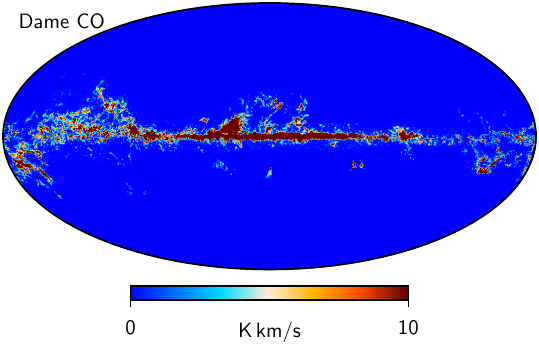}
  \caption{Input maps of each of the following thermal dust tracers (from left to right and top to bottom): \CII\ 158\,$\mu$m line emission from \COBE-FIRAS, \HI\ line emission from HI4PI, dust extinction from \GAIA, H$\alpha$ line emission from WHAM, and CO line emission from \citet{dame:2001}. All maps are smoothed to the FIRAS beam and then used for the low-resolution template fitting at \nside=16.}
  \label{fig:tempfit_inputs}
\end{figure*}

\begin{figure}
  \centering
  \includegraphics[width=\columnwidth]{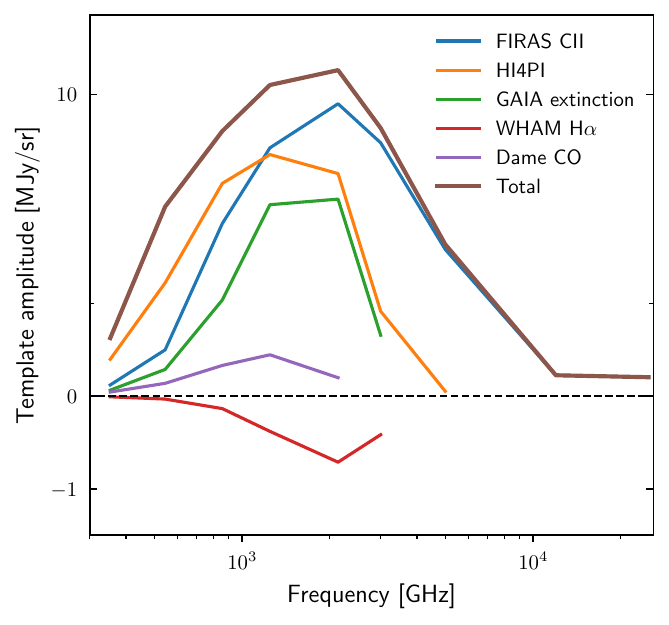}
  \caption{Template amplitudes as a function of frequency for the nine channels used in this analysis, plotted only at the channels where they are fit. The WHAM H$\alpha$ amplitudes are negative because it is an absorptive template.}
  \label{fig:tempamp}
\end{figure}

\begin{figure}
  \centering
  \includegraphics[width=\columnwidth]{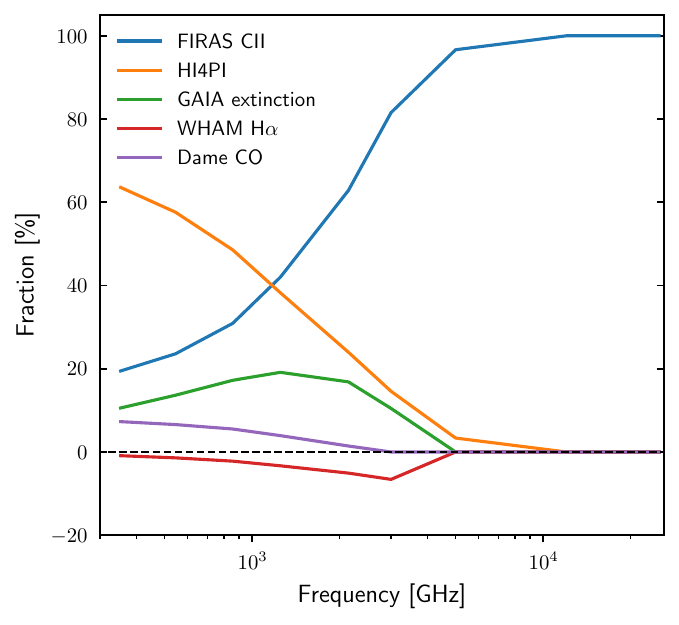}
  \caption{Relative template amplitudes at each frequency in terms of the power contribution of each component as a percentage of the combined model amplitude.}
  \label{fig:tempfract}
\end{figure}

\begin{figure*}
  \centering
  \includegraphics[width=0.40\textwidth]{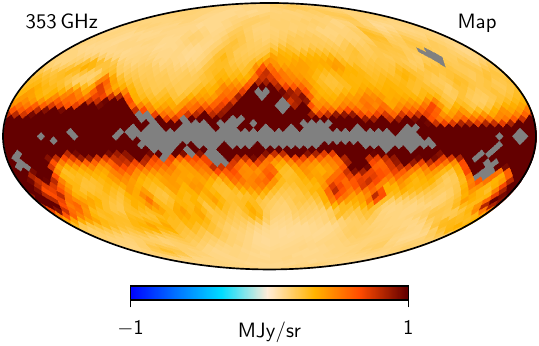}
  \includegraphics[width=0.40\textwidth]{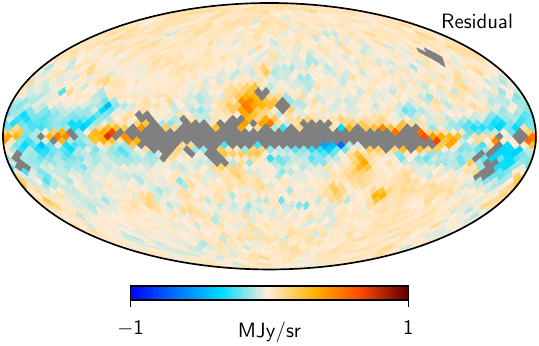}\\
  \includegraphics[width=0.40\textwidth]{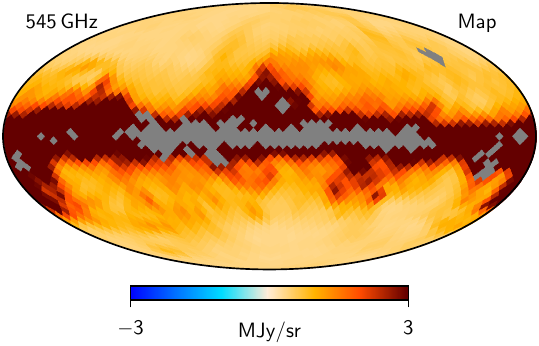}
  \includegraphics[width=0.40\textwidth]{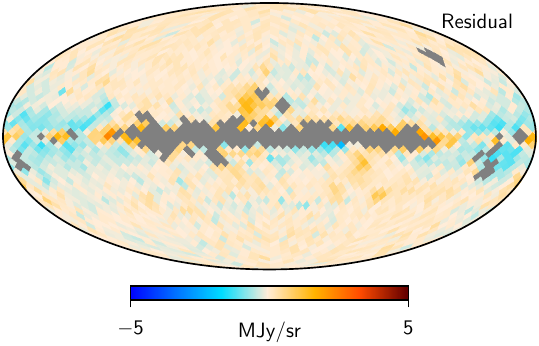}\\
  \includegraphics[width=0.40\textwidth]{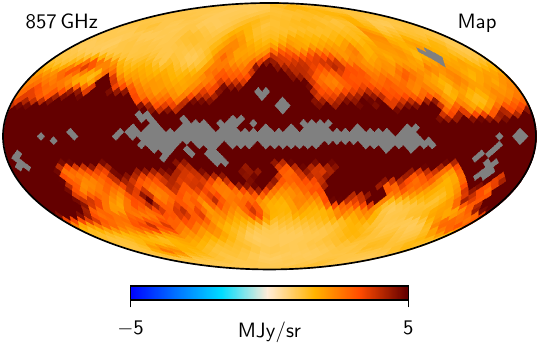}
  \includegraphics[width=0.40\textwidth]{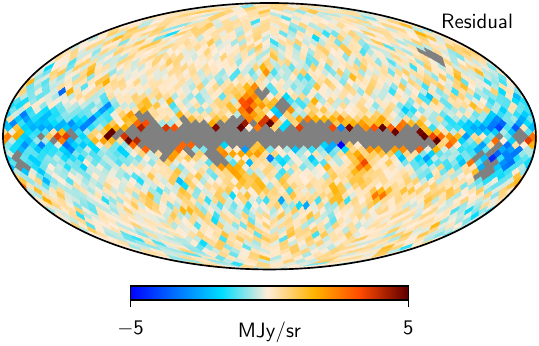}\\
  \includegraphics[width=0.40\textwidth]{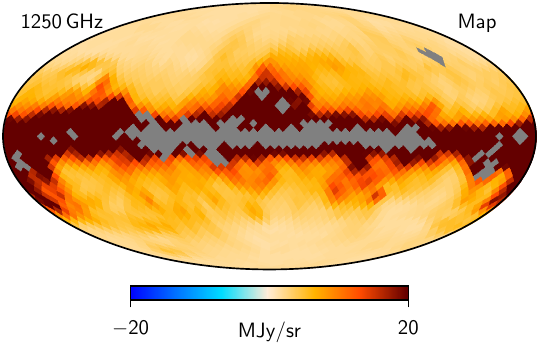}
  \includegraphics[width=0.40\textwidth]{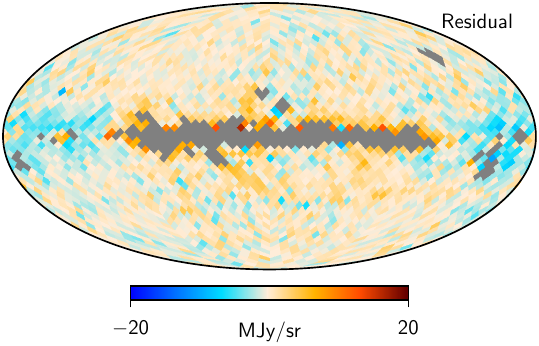}\\  
  \includegraphics[width=0.40\textwidth]{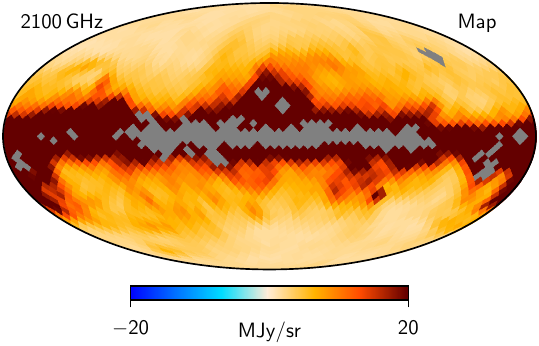}
  \includegraphics[width=0.40\textwidth]{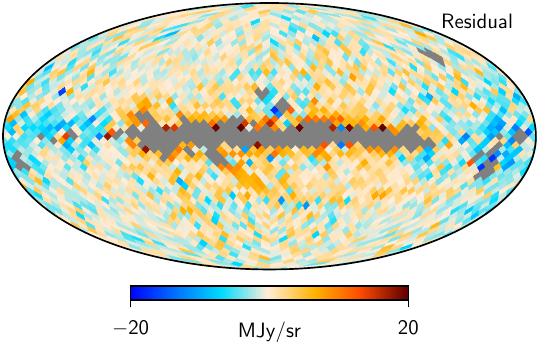}
  \caption{Comparison between full (left column) and template cleaned
    (right) maps between 353 and 2100\,GHz. The grey pixels are masked
    by the common analysis mask. All maps are convolved with the FIRAS
    beam. }
  \label{fig:tempfit_map_vs_res1}
\end{figure*}

\begin{figure*}
  \centering
  \includegraphics[width=0.40\textwidth]{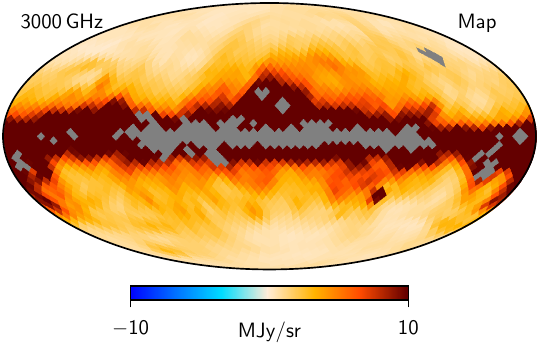}
  \includegraphics[width=0.40\textwidth]{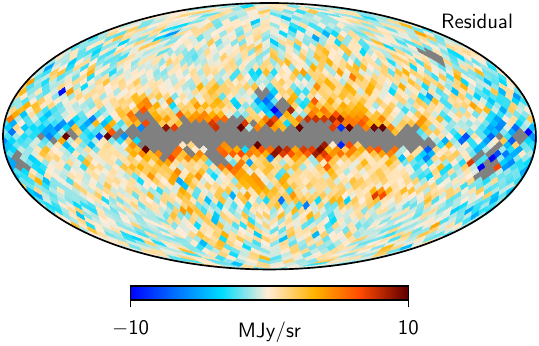}\\
  \includegraphics[width=0.40\textwidth]{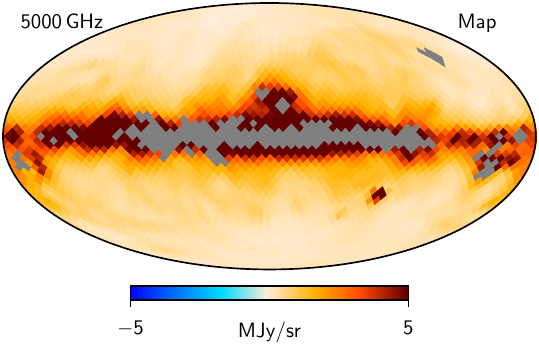}
  \includegraphics[width=0.40\textwidth]{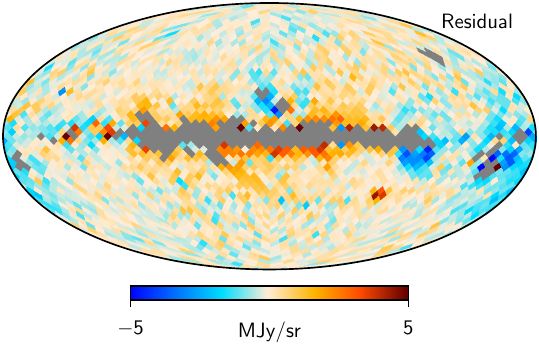}\\
  \includegraphics[width=0.40\textwidth]{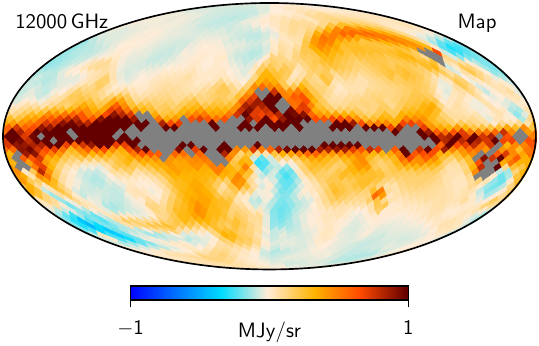}
  \includegraphics[width=0.40\textwidth]{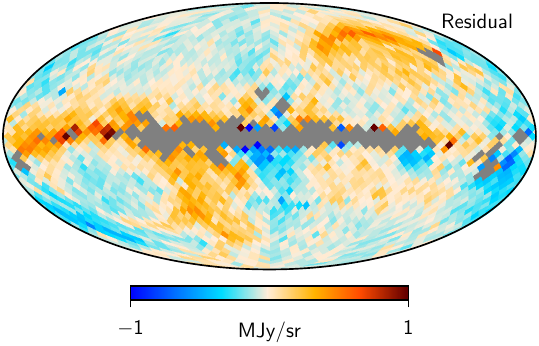}\\
  \includegraphics[width=0.40\textwidth]{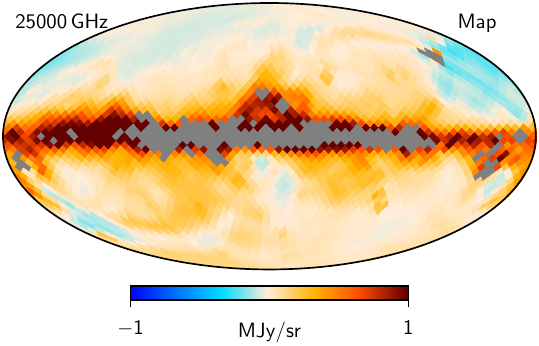}
  \includegraphics[width=0.40\textwidth]{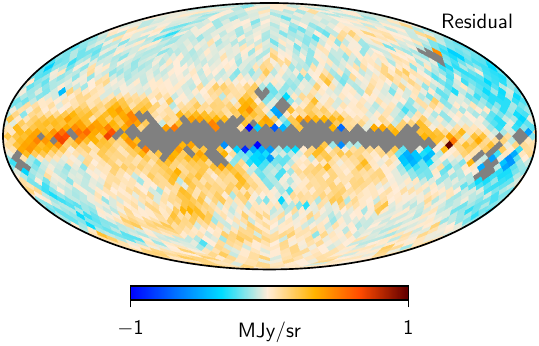}
  \caption{Same as \cref{fig:tempfit_map_vs_res1} but frequencies
    between 3000 and 25\,000\,GHz (100\,$\mu$m - 12\,$\mu$m).}
  \label{fig:tempfit_map_vs_res3}
\end{figure*}

\begin{figure}
  \centering
  \includegraphics[width=\columnwidth]{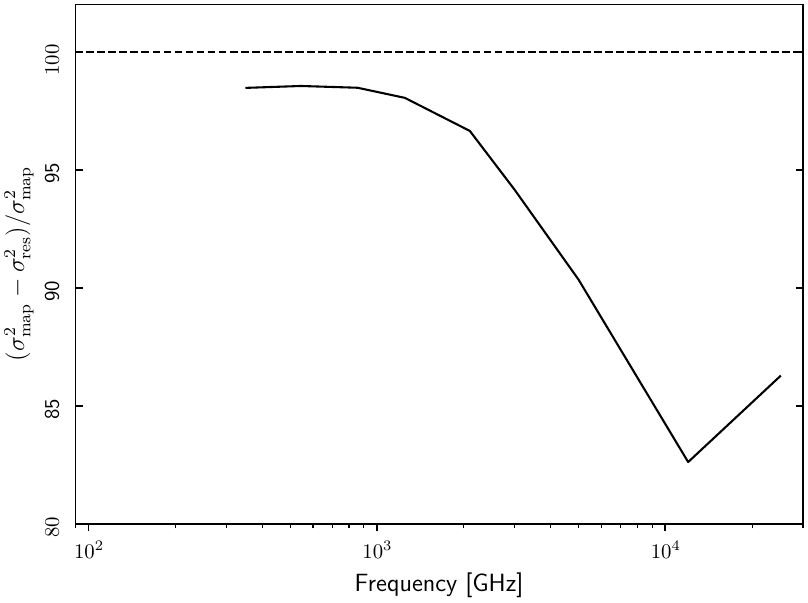}
  \caption{Efficiency of the dust model at describing the data at at each frequency. We subtract the squared residuals from the squared map, and normalize by the amplitude of the map. The five component template-based dust model is able to recover more than 80\% of the total map power at all frequencies, including those with known residual zodiacal contamination like 25\,$\mu$m.}
  \label{fig:temprms}
\end{figure}

Next, we define $\a = \{a_c,m\}$ to be the vector of all free template
and monopole amplitudes, for a total of six parameters per frequency
channel, and we assume the noise to be Gaussian distributed with a
covariance matrix $\N_i$. Fitting a linear model to Gaussian
data has a well-known solution, and reads
\begin{equation}
    \boldsymbol{a}_i = (\mathsf{T}^T \N^{-1} \mathsf{T})^{-1} \mathsf{T}^T\N^{-1}\boldsymbol{d},
    \label{eq:lss_full}
\end{equation}
where we define $\mathsf{T}$ to be an $n_\mathrm{pix} \times
(n_{\mathrm{comp}}+1)$ template matrix such that matrix column $c$
contains component template $\t_c$ and the final column
contains the constant unity for the monopole. This equation is easily
solved brute-force by matrix inversion.

The noise covariance, $\N$, should ideally describe all sources of
uncertainty, both systematic uncertainties and random noise. However,
since we are working with a very low-dimensional data model fitted to
datasets with extremely high signal-to-noise ratios, the random
instrumental noise is negligible compared to calibration and
modelling uncertainties, and we therefore set $\N$ equal to a constant
in the following. \Cref{eq:lss_full} therefore simplifies to 
\begin{equation}
    \boldsymbol{a}_i = (\mathsf{T}^T \mathsf{T})^{-1} \mathsf{T}^T\boldsymbol{d},
    \label{eq:lss}
\end{equation}
which is the standard normal equation for uniform weights.

Masking of bright or unobserved pixels is implemented simply by
removing the relevant pixels in $\t$ and
$\boldsymbol{d}$. Specifically, we remove pixels very close to the
Galactic plane, as well as any pixels excluded by the FIRAS processing
mask. A total of 91\,\% of the sky is included the following analysis.

\section{Data}
\label{sec:data}

\subsection{Frequency maps}

In this paper we consider only data for which thermal dust dominates
strongly over other astrophysical contributions. Specifically, we
consider the \Planck\ High Frequency Instrument (HFI) 353, 545, and
857 GHz frequency maps \citep{planck2016-l03} and the six longest
wavelength DIRBE bands from 240\,$\mu\mathrm m$ to $12\,\mu\mathrm m$.

For HFI, we use \Planck\ PR4 \citep{npipe} frequency sky maps. From
these, we subtract 1) the best-fit CMB dipole listed in the PR4 paper;
2) CMB fluctuations as estimated by the SMICA algorithm
\citep{planck2016-l04}; 3) CIB fluctuations as estimated by the GNILC
algorithm \citep{planck2016-XLVIII}; and 4) an estimate of the
stationary zodiacal light emission \citep{planck2013-pip88}. We have
purposefully chosen to use products not generated by \Commander, in
order for this paper to be as independent of the main \Cosmoglobe\ DR2
analysis as possible.

Ideally, we would have preferred to use products that are independent
of \Cosmoglobe\ DR2 also for DIRBE, but in this case the official
products are too contaminated by zodiacal light residuals for our
purposes. We therefore use the DIRBE frequency maps presented by
\citet{CG02_01}. For these, no additional
contributions are subtracted.

Since this work is primarily concerned with dust modeling, we attempt to avoid
contamination from (especially) the CMB signal, which dominates at the
lower \Planck\ frequencies. To that end, we use only the HFI data from
353 to 857\,GHz. All of the input frequency maps are taken from the \Planck\ PR4 data release
\citep{npipe}.

The template fitting procedure described in \cref{sec:tempfit}
requires all maps to be smoothed to a common resolution -- if not, the
extra degrees of freedom at smaller scales in some of the maps and
templates might bias the fit at larger scales. Out of all the data and
templates used in this analysis, the \COBE-FIRAS \CII\ map has the lowest native resolution. We therefore convolve all involved
maps (except \CII) with the \COBE-FIRAS beam, shown in \cref{fig:firasbeam} (taking
into account the drift along Ecliptic meridians; e.g., \citealp{odegard:2019}), and repixelating to a common HEALPix
resolution of \nside=16.

\subsection{Thermal dust templates}

To model the thermal dust contribution in HFI and DIRBE, we consider
the five following dust tracers, plotted at full angular resolution in Fig.~\ref{fig:tempfit_inputs}.

\paragraph{Ionized carbon --- \COBE-FIRAS \CII}
The \COBE-FIRAS instrument \citep{mather:1994} is the only currently
available full-sky spectrometric survey in the microwave and infrared
electromagnetic wavelengths. While the original science goal of the
experiment was to measure the CMB monopole spectrum, it also provided
a wealth of large-scale Galactic information as well. In particular,
it provided the best full-sky measurement of Galactic
$157.7\,\mu\mathrm m$ \CII\ emission published to date
\citep{fixsen:1998}. \CII\ plays a key role in Galactic structure
formation as a key cooling agent, and it has long been
recognized as a powerful gas tracer; in the current paper, however, we
consider its potential as a dust tracer. 

\paragraph{Neutral hydrogen -- HI4PI}
The HI4PI HI survey \citep{HI4PI:2016} is a full-sky survey of the
21\,cm hydrogen line, performed by a combination of data from EBHIS
and GASS. The 21 cm line has been used as a tracer for dust structures
\citep{planck2011-7.12}, and is typically associated with
low-intensity regions of the sky, as it requires the hydrogen to not
be ionized, but rather be in its lowest energy state.

\paragraph{Nearby dust through extinction -- Gaia}
One source of important information about the 3-dimensional structure
of dust is starlight extinction. By comparing expected stellar spectra
with observed ones, \cite{edenhofer:2024} was able to reconstruct such
dust structures up to a distance of 1.25\,kpc, represented by a series
of maps for each distance bin. Using the accompanying software, we
integrated these maps out to their furthest distance, which still
represents relatively close dust structures, and we use this as a
template for nearby dust structures. The choice of the distance cut is
driven by the available Edenhofer maps rather than any physical
considerations.

\paragraph{Ionized hydrogen -- WHAM}
The Wisconsin H$\alpha$ Mapper (WHAM) \citep{wham_north:2003,
  wham_south:2010} surveyed the intensity of the H$\alpha$ line over
the whole sky, with a resolution of $1^{\circ}$ FWHM and a velocity
resolution of 12\,km/s. This line is typically associated with higher
energy regions, as it is most commonly emitted from hydrogen atoms
recombining after ionization. The WHAM H$\alpha$ map was used in early
\Planck\ analyses as a tracer of nearby free-free emission.

\paragraph{Carbon monoxide -- Dame CO\,$J=1$-0}
Finally, \citet{dame:2001} surveyed about 45\% of the sky for the
$J$=$1\rightarrow0$ CO line within 30 degrees of the Galactic
equator. CO is a well-known tracer of cold dust surrounding star
forming regions.

It is important to note that the relative importance of these tracers
depend strongly on both frequency and position on the sky. When
performing the first iteration of the analysis presented in the
following, we made no assumptions regarding the relative contribution
of each component, but simply fitted all five components freely to all
channels. However, due to the fact that the set of five selected
templates discussed here certainly represents an incomplete basis for
the true dust signal, the fitted amplitudes for a given component tend
to become unphysical when their signal-to-noise ratio approaches zero;
rather than fitting an actual component, the fitting algorithm uses it
to suppress modelling errors in the others, and the result is
typically obvious ``over-subtraction shadows'' in the data-minus-model
residual maps. For this reason, we exclude individual templates from
the fit at the higher frequencies, once we observe through inspection
of the resulting spectra and residuals that the component in question
does not significantly improve the overall fit. \cref{tab:bands}
provides an overview of which templates are fitted to which frequency
maps.

\section{Results}
\label{sec:results}

\subsection{Template amplitude fits}
\label{sec:tempamp}

The template amplitudes per frequency band resulting from applying the
algorithm in \cref{sec:tempfit} are shown in \cref{fig:tempamp}. The
thick brown curves shows the sum of all components. First, we note
that all spectra show a characteristic bump with a peak frequency
between $\sim$1200 and 2000\,GHz, which corresponds to MBB
temperatures between 10 and 30\,K; see below for actual temperature
estimations of these spectra.

Second, at frequencies below $\sim$1200\,GHz the overall fit is
dominated by HI4PI, which traces neutral hydrogen. This agrees well
with our prior expectations, given the wide usage of HI as a dust
tracer in the microwave regime
\citep[e.g.,][]{planck2011-6.6,lenz:2017,Hensley2023}.

However, it is perhaps more surprising to note that the total fit
above 1200\,GHz is dominated by the FIRAS \CII\ map. Indeed, above
3000\,GHz \CII\ is effectively the only tracer with a statistically
significant contribution. This simple yet far-reaching observation
represents the main finding in this paper.

Another potentially surprising result is that the H$\alpha$ component
appears with negative amplitudes in Fig.~\ref{fig:tempamp}. This
indicates that regions dominated by H$\alpha$ acts as a dust
extinction source in the current model, rather than an emission
source. We have spent significant time trying alternative models
and data configurations, but have found that this observation is
extremely robust against data selection; whenever the H$\alpha$ map is
excluded, a blue imprint with a H$\alpha$-like morphology appears in
the residuals. 

A complementary view of these spectra is provided in
\cref{fig:tempfract}, which shows the relative fraction of each
component relative to the sum as a function of frequency. More than
60\,\% of the model is described by the HI component at 353\,GHz,
while at 1000\,GHz the HI and \CII\ contributes about 40\,\% each,
with the nearby \Gaia\ dust map accounting for most the remaining
20\,\%. The H$\alpha$ and CO tracers account for roughly comparable
quantities, but with opposite signs. Once again, we see that above
3000\,GHz the \CII\ component fully dominates.

\subsection{Model efficiency and goodness-of-fit}
\label{sec:residuals}
In
\cref{fig:tempfit_map_vs_res1,fig:tempfit_map_vs_res3},
we show a comparison between the original data (left columns) and the
template cleaned maps (right columns). Clearly, the template fitting
performs well for all frequencies, leaving residuals that are much
smaller in amplitude than the original maps. The morphologies of those
residuals appear to correlate with high-activity regions of the Milky
Way, in particular at 857\,GHz and 3000\,GHz, as well as unmodelled
zodiacal emission from the Solar System, which is most evident in the
two highest frequency maps.

In order to quantify the degree to which these visual impressions
bears out in practice, we show in \cref{fig:temprms} a measure of the
amount of signal that is accounted for by the template fit. This
measure is simply defined by the ratio between the variance difference
between the raw data and the residual and the variance of the raw data
itself. If the model happened to fully model the data, the residual
variance would be zero, and the ratio would be unity; if the model
happened to explain nothing of the data, the residual would be equal
to the data, and the ratio would be zero. In Fig.~\ref{fig:temprms} we
see that this quantity ranges between 98\,\% at frequencies below
1000\,GHz, and decreases to about 83\,\% at 12\,000\,GHz. In this
respect, it is important to note that zodiacal light residuals become
gradually stronger with frequency, and they are particularly important
at the DIRBE 25 and 12\,$\mu$m channels.

It is also worth emphasizing that the only active dust tracer at
12\,$\mu$m is \CII\ (see Table~\ref{tab:bands}), and still the
efficiency is higher than 85\,\%. This clearly demonstrates the
power of the \CII\ map as a high-frequency dust tracer. 

\subsection{Modified blackbody fits}
\label{sec:mbb}
\begin{figure}
  \centering
  \includegraphics[width=\columnwidth]{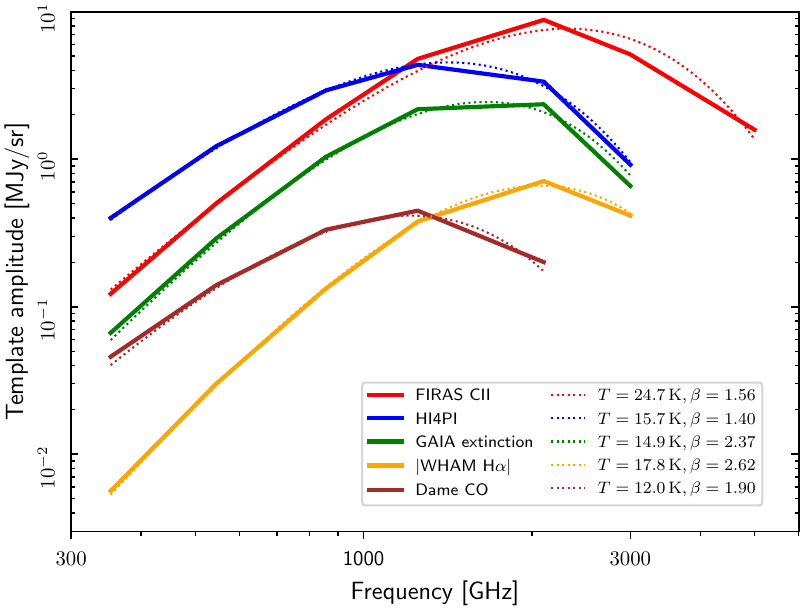}
  \caption{Comparison between template fit amplitudes (solid lines,
    reproduced from \cref{fig:tempamp}) with best-fit modified
    blackbody models (dotted lines), in the thermal regime for each
    component. The best-fit MBB temperatures and spectral indices for each curve are listed in the figure legend.}
  \label{fig:Bnu_comparison}
\end{figure}

In \cref{fig:Bnu_comparison}, we fit a modified blackbody model
to the template fitting SEDs,
\begin{equation}
    s(\nu) = A \frac{\nu^{\beta + 3}}{\exp{(\frac{h\nu}{k_b T})} - 1}.
\end{equation}
Here, $A$ is an arbitrary scaling factor, $T$ is the blackbody
temperature, and $\beta$ is the spectral index. It is especially
interesting to compare the two strongest SEDs --- those associated
with the \CII\ and \HI\ templates, respectively --- as we see that they
correspond to significantly different temperatures. The \HI-correlated
component has a best-fit temperature of about 16\,K, while the
\CII-correlated component has a best-fit temperature of about
25\,K. The corresponding best-fit spectral indices are $\beta=1.40$
and $1.56$, respectively. For this reason, we refer to the \HI- and
\CII-correlated components as ``cold'' and ``hot dust'' in our
companion papers, although we note that the latter is often used about
significantly higher temperatures in the literature.

The CO-correlated component corresponds to the lowest temperature,
around 12\,K. This component is known to trace cold and dense
star-forming regions, and a lower temperature is therefore in good
agreement with prior expectations.

Moving on to the nearby \GAIA\ extinction template, we also find a
somewhat low temperature, around 15\,K, which is similar to the \HI\
temperature. However, we see that the actual SED peaks between the \HI\
and \CII\ SEDs, which is due to a very steep spectral index. In this
respect, it is worth recalling that this nearby component has a lower
amplitude, and therefore lower signal-to-noise ratio, than the two
dominant components, and it is therefore more susceptible to modelling
errors and data selection effects. For comparison, when fitting these
spectral parameters to HFI data, \cite{CG02_06} finds a best-fit MBB
temperature for the nearby component of 18\,K.

\section{Conclusions}
\label{sec:conclusions}

In this paper, we have fit five pre-existing dust tracers to ten
\Planck\ HFI and DIRBE frequency maps between 353\,GHz and
25\,THz. These include a FIRAS \CII\ template; an \HI\ HI4PI template;
a ``nearby dust'' template constructed from integrated
\GAIA\ extinction maps; the Dame CO template; and a WHAM H$\alpha$
template. We find that this simple template-based model is both able
to account for more than 95\% of the dust variance for all but the two
highest frequency bands, and that the resulting SEDs follow physically
reasonable modified blackbody curves up to about 3-5\,THz.

The most important novel finding in this analysis is that the most
efficient tracer of large-scale dust emission is \CII. The importance
of \CII\ emission for Galactic structure, dynamics and cooling physics
in general has been long recognized, but the fact that it also is an
extremely efficient dust tracer is to our knowledge a new result. In
fact, above 3000\,GHz this component totally dominates the overall
fit, and even at \Planck\ HFI frequencies it is second only to \HI. The
temperature of this \CII-correlated component is significantly higher
than for either the \HI\ or nearby components, with a typical value of
about 25\,K. 

This analysis indicates that it is possible to vastly reduce the
complexity of the challenge of thermal dust modelling for both
microwave and infrared regimes. Instead of operating with
phenomenological models that involve multiple free parameters per
pixel, a model based on multiple physical components, each with
well-defined spectral parameters, could both lead to a lower number of
degrees of freedom in total, and also provide a more direct way of
imposing physically well-motivated priors and combining data
sets. This general approach is applied in practice in two companion
papers by \citet{CG02_06} and \citet{CG02_07}, who fit multi-component
dust models to the full-resolution \Planck\ HFI and DIRBE data,
respectively. 

A final key question arising from this analysis is whether these
findings also apply to polarized dust emission, and this question will
be the subject of a future analysis. If this turns out to be the case,
it would have a significant impact on the feasibility of modelling
polarized dust emission for future CMB experiments, which are
dependent on accurate modelling of polarized foregrounds to reach the
sensitivities required for their science goals (in particular, the
detection of tensor-to-scalar ratio $r$). In particular, a key goal of future CMB $B$-mode
experiments would then be to internally distinguish between hot and
cold dust, which in turn would impose requirements on the frequency
coverage of the experiment in question, with an increased focus on
frequencies between 500 and 1000\,GHz.

\begin{acknowledgements}
  We thank Richard Arendt, Tony Banday, Johannes Eskilt, Dale Fixsen, Ken Ganga, Paul
  Goldsmith, Shuji Matsuura, Sven Wedemeyer, Janet Weiland and Edward Wright for
  useful suggestions and guidance.  The current work has received
  funding from the European Union’s Horizon 2020 research and
  innovation programme under grant agreement numbers 819478 (ERC;
  \textsc{Cosmoglobe}), 772253 (ERC; \textsc{bits2cosmology}),
  101165647 (ERC, \textsc{Origins}), 101141621 (ERC,
  \textsc{Commander}), and 101007633 (MSCA; \textsc{CMBInflate}).
  This article reflects the views of the authors only. The funding
  body is not responsible for any use that may be made of the
  information contained therein. This research is
  also funded by the Research Council of Norway under grant agreement
  numbers 344934 (YRT; \textsc{CosmoglobeHD}) and 351037 (FRIPRO;
  \textsc{LiteBIRD-Norway}). Some of the results in this paper have been
  derived using healpy \citep{Zonca2019} and the HEALPix
  \citep{healpix} packages.  We acknowledge the use of the Legacy
  Archive for Microwave Background Data Analysis (LAMBDA), part of the
  High Energy Astrophysics Science Archive Center
  (HEASARC). HEASARC/LAMBDA is a service of the Astrophysics Science
  Division at the NASA Goddard Space Flight Center. This publication
  makes use of data products from the Wide-field Infrared Survey
  Explorer, which is a joint project of the University of California,
  Los Angeles, and the Jet Propulsion Laboratory/California Institute
  of Technology, funded by the National Aeronautics and Space
  Administration. This work has made use of data from the European
  Space Agency (ESA) mission {\it Gaia}
  (\url{https://www.cosmos.esa.int/gaia}), processed by the {\it Gaia}
  Data Processing and Analysis Consortium (DPAC,
  \url{https://www.cosmos.esa.int/web/gaia/dpac/consortium}). Funding
  for the DPAC has been provided by national institutions, in
  particular the institutions participating in the {\it Gaia}
  Multilateral Agreement.  We acknowledge the use of data provided by
  the Centre d'Analyse de Données Etendues (CADE), a service of
  IRAP-UPS/CNRS (http://cade.irap.omp.eu, \citealt{paradis:2012}).  
  This paper and related research have been conducted during and with the support of the Italian national inter-university PhD programme in Space Science and Technology. Work on this article was produced while attending the PhD program in PhD in Space Science and Technology at the University of Trento, Cycle XXXIX, with the support of a scholarship financed by the Ministerial Decree no. 118 of 2nd March 2023, based on the NRRP - funded by the European Union - NextGenerationEU - Mission 4 "Education and Research", Component 1 "Enhancement of the offer of educational services: from nurseries to universities” - Investment 4.1 “Extension of the number of research doctorates and innovative doctorates for public administration and cultural heritage” - CUP E66E23000110001.
\end{acknowledgements}

%-------------------------------------------------------------
%                                       Table with references 
%-------------------------------------------------------------
%

\bibliographystyle{aa}
\bibliography{../../common/CG_bibliography,references,../../common/Planck_bib}
\end{document}